# Nature of the positron state in CdSe quantum dots

Wenqin Shi[1,*], Vincent Callewaert[2,*], Bernardo Barbiellini[3,4], Rolando Saniz[2], Maik Butterling[1], Werner Egger[5], Marcel Dickmann[5], Christoph Hugenschmidt[6], Behtash Shakeri[7], Robert W. Meulenberg[7], Ekkes Brück[1], Bart Partoens[2], Arun Bansil[4], and Stephan W.H. Eijt[1,†]

[1]*Fundamental Aspects of Materials and Energy, Faculty of Applied Sciences, Delft University of Technology, Mekelweg 15, NL-2629 JB Delft, Netherlands*
[2]*Department of Physics, University of Antwerp, B-2020 Antwerp, Belgium*
[3]*School of Engineering Science, Lappeenranta University of Technology, FI-53851 Lappeenranta, Finland*
[4]*Department of Physics, Northeastern University, Boston, MA 02115, USA*
[5]*Institut für Angewandte Physik und Messtechnik, Universität der Bundeswehr München, D-85579 Neubiberg, Germany*
[6]*Physics Department & Heinz Maier-Leibnitz Zentrum (MLZ), Technische Universität München, D-85748 Garching, Germany*
[7]*Department of Physics and Astronomy and the Laboratory for Surface Science and Technology, University of Maine, Orono ME 04469, USA*

Date: *May 30, 2018*

Previous studies have shown that positron-annihilation spectroscopy is a highly sensitive probe of the electronic structure and surface composition of ligand-capped semiconductor Quantum Dots (QDs) embedded in thin films. Nature of the associated positron state, however, whether the positron is confined inside the QDs or localized at their surfaces, has so far remained unresolved. Our positron-annihilation lifetime spectroscopy (PALS) studies of CdSe QDs reveal the presence of a strong lifetime component in the narrow range of 358-371 ps, indicating abundant trapping and annihilation of positrons at the surfaces of the QDs. Furthermore, our ab-initio calculations of the positron wave function and lifetime employing a recent formulation of the Weighted Density Approximation (WDA) demonstrate the presence of a positron surface state and predict positron lifetimes close to experimental values. Our study thus resolves the longstanding question regarding the nature of the positron state in semiconductor QDs, and opens the way to extract quantitative information on surface composition and ligand-surface interactions of colloidal semiconductor QDs through highly sensitive positron-annihilation techniques.

PhySH: Quantum Dots, Positron Annihilation Spectroscopy, Ab initio Calculations, Surface States, Vacancies, Surfaces, Solar Cells, Thin Films, LDA

Colloidal semiconductor Quantum Dots (QDs) are drawing intense interest as potential functional building blocks for next-generation photovoltaic (PV) devices due to their special size-tunable optoelectronic properties. Solar cells based on PbS QDs recently reached promising efficiencies above 10% [1]. Relative to the bulk solid, QDs possess a very high surface-to-volume ratio, which greatly magnifies contributions of the surface structure, composition and electronic structure to their properties [2-4]. In order to prevent defect states in the band gap from dangling bonds of under-coordinated surface atoms, it is important to passivate such surface states via suitable ligand molecules, since imperfect passivation can severely limit the efficiency of QD solar cells [5]. For boosting the performance of QD-based PV devices, a deeper understanding of the electronic and surface structure of QDs capped with surface ligands is necessary. In this connection, recent years have witnessed innovative applications of X-ray absorption and positron-annihilation techniques [6-11] and the development of a wide range of computational modeling methods [12-15].

Studies have shown that positron techniques are effective probes of surface composition and electronic structure of semiconductor QDs [9, 10, 16, 17]. However, the cause of the high surface sensitivity, and a firm theoretical understanding of the underlying positron state is still lacking. The first experimental positron studies of colloidal CdSe QDs by Weber *et al*. [9] demonstrated that positron Coincidence Doppler Broadening can sensitively probe the electronic structure of semiconductor QDs via their positron-electron momentum distributions (PEMDs), attributed to positron annihilation from a positron state confined inside of the QD, that has characteristics of a confined 'bulk'-like state. Later, Eijt *et al*. [10] revealed that the positron is mainly located at the surfaces of CdSe QDs, facilitated by the comparison of the PEMDs of CdSe QDs and bulk CdSe observed by positron 2D-ACAR (two-dimensional Angular-Correlation-of-Annihilation-Radiation) with the PEMD of bulk CdSe obtained from *ab-initio* calculations. The positron wave function was only schematically described as a 'shell'-like state at the surface of the QD. Notably, recent Positron Annihilation Lifetime Spectroscopy (PALS) measurements indicated the presence of a positron surface state for PbSe QDs as well [16].

In order to provide evidence of the positron surface state in CdSe QDs indicated by positron experiments, *ab-initio* calculations of the positron-surface interaction potential and the resulting positron wave function are essential. The positron wave function is subsequently used to calculate the annihilation probabilities, enabling direct comparison with PALS experiments. Positron states at solid surfaces were extensively studied in the past years [18-20]. However, unlike the case of a positron in a bulk solid, the screening cloud of electrons at the surface is strongly anisotropic and the positron correlation potential cannot be accurately described by the local-density approximation (LDA), which has proven to be reliable for the description of the bulk positron wave function in solids [21]. This has prevented a satisfactory theoretical treatment of the positron wave function at the surfaces of QDs, even in studies employing a combined LDA plus corrugated mirror model (CMM) approach, where the erroneous behavior of the LDA in the vacuum region is empirically corrected.

In this Letter, we demonstrate the existence of a positron surface state in CdSe QDs through quantitative and systematic comparison of PALS spectra with corresponding first-principles calculations. Two methods, namely, the LDA+CMM [22, 23] and the recently developed implementation of the WDA [24, 25], are used to model the electron-positron correlation potential. In contrast to the standard implementation of the LDA+CMM, our calculations within the WDA show that the positron wave function is characterized consistently by the presence of a clear peak located at the surface of CdSe QDs, with calculated lifetimes close to the experimental values.

CdSe QDs with a mean diameter of 6.5 nm were produced using the synthesis methods described in Ref. [26], which produces QDs nominally capped with both stearic acid (SA) and trioctylphosphine oxide (TOPO) ligands. In Ref. [26], we showed that the type of ligand present on the surface of the QD can be controlled during post-processing, so that only SA is left on the surface. Ligand exchange was then performed with three ligand types: oleylamine (OLA), oleic acid (OA), and TOPO. In this way, five unique samples were obtained: CdSe coated with SA/TOPO, SA, TOPO, OLA, and OA. CdSe QD layers with thicknesses in the range of several μm were produced by drop-casting of the solutions on 1×1 $cm^2$ ITO-coated glass substrates.

These thin-film samples were examined by PALS [27, 28, 19] using the pulsed low-energy positron lifetime spectrometer (PLEPS) instrument [29] of the neutron induced positron source (NEPOMUC) facility [30] at the Heinz Maier–Leibnitz Zentrum (MLZ) research reactor in Garching. Measurements were performed at selected positron energies between 1 and 18 keV. Around 4·$10^6$ counts were collected for each lifetime spectrum. The lifetime spectra were fitted by using the LT program [31].

Representative PALS spectra are shown in Figure 1(a) for 6.5-nm CdSe QDs. The spectra are remarkably similar in the short-time region, showing only differences upon variation of the binding ligand in mainly the intensity of the long-lifetime component characteristic of ortho-positronium (o-Ps) formation. Satisfactory lifetime fits could be obtained using a three component decomposition, as exemplified in Figure 1(b). The three components correspond to: (1) para-positronium (p-Ps), (2) a positron lifetime of 370 ps, which is in the range of values indicating positron surface annihilation [16, 32], and (3) ortho-positronium. In the fit, the instrumental resolution function is described by a sum of two Gaussian functions, obtained by measurement of the PALS spectra of p-doped SiC sample for each investigated positron implantation energy, assuming a bulk p-SiC lifetime of 145 ps and surface lifetime of 385 ps [33]. In the analysis of the PALS spectra, a constraint of $I_{o-Ps} = 3I_{p-Ps}$ due to the spin multiplicity of the positronium states [34] was applied. Table 1 lists the fitted parameters obtained, which do not show any significant dependence on the positron implantation energy beyond 2 keV, indicating the homogeneity of the QD layers.

Table 1. Positron lifetimes and intensities for CdSe QDs capped with five different ligands averaged over values obtained for five positron implantation energies in the range of 2-16 keV. Errors are standard deviations in the average values. Last column gives the range of fit variances.

| Capping ligand | $I_1$ (%) | $I_2$ (%) | $I_3$ (%) | $\tau_1$ (ps) | $\tau_2$ (ps) | $\tau_3$ (ns) | Fit variance range |
|---|---|---|---|---|---|---|---|
| OLA | 5.7±0.2 | 77±1 | 17.0±0.5 | 101±2 | 371±1 | 2.93±0.05 | 1.16 – 1.27 |
| OA | 5.4±0.8 | 78±3 | 16.2±2.5 | 101±13 | 363±1 | 3.29±0.15 | 1.33 – 1.55 |
| TOPO | 4.4±0.6 | 82±2 | 13.3±1.7 | 110±5 | 369±3 | 2.83±0.04 | 1.19 – 1.30 |
| SA/TOPO | 3.6±0.8 | 86±3 | 10.7±2.4 | 92±6 | 358±2 | 2.85±0.04 | 1.49 – 1.86 |
| SA | 3.9±0.9 | 84±4 | 11.7±2.7 | 100±22 | 364±2 | 2.84±0.04 | 1.38 – 1.69 |

The PALS study reveals the presence of a dominant (77%-86%) second lifetime component with lifetimes in the narrow range of 358-370 ps, *i.e.* significantly higher than the experimental positron lifetime of defect-free CdSe (275 ps, [9]), and at most weakly depending on the type of ligand present. The longest lifetime component is in the range of 2.8-3.3 ns with an intensity of 10-17%, corresponding to o-Ps that forms in the open spaces between the carbon chains of surface ligands or at the surface of the QDs and annihilates via pick-off annihilation [16]. The o-Ps lifetimes indicate an open space size of about 0.7-0.8 nm [35, 36]. The relative intensities of the second lifetime component and Ps are comparable to the estimated fractions of positrons stopped in the CdSe core and ligand shells, 88% and 12%, respectively, extracted from the mass-density-weighted volume fractions for the CdSe cores and ligand shells [16]. This provides further indication that Ps is formed in the ligands, while it strongly suggests that the intense second lifetime component arises mainly from the majority of positrons that are stopped and thermalize in the QD cores. In view of the small size of the QDs, the wave function of these positrons will have considerable overlap with the surfaces of the QDs and these positrons may thus easily trap in a surface state. The transition from a 'bulk'-like state confined in the QD to a surface state in the last processes of thermalization may occur for example via an Auger process [20].

Clearly, a satisfactory and robust decomposition of the PALS spectra into three components was obtained for all samples, with the shortest and longest lifetime component corresponding to p-Ps and o-Ps (pick-off) annihilation, respectively, and a dominant intermediate lifetime associated with positron surface state annihilation. In the analysis, the constraint of $I_{o-Ps} = 3I_{p-Ps}$ leads to p-Ps lifetimes of about 90-110 ps, *i.e.* close to the intrinsic (vacuum) lifetime of 125 ps. It should be noted that the time resolution (ranging from 260 ps to 280 ps) was not optimal and could be a source of uncertainty in the quantitative determination of the (short) p-Ps lifetime. We checked the reliability and robustness of the PALS analysis by comparison to another fitting scheme in which the lifetime of p-Ps is fixed at its vacuum value of 125 ps. In that case also, satisfactory fits can be obtained, but they lead to unphysically large deviations from the expected 1:3 ratio in p-Ps to o-Ps intensities in absence of strong magnetic fields. Importantly, however, the lifetimes and intensities of the second lifetime component remain nearly the same as in Table 1, demonstrating the robustness of the lifetime parameters characterizing the positron surface state.

In order to provide firm support for the positron surface state at the surfaces of QDs inferred from the PALS experiments, we have performed first-principles calculations of the positron ground state wave function and the corresponding positron annihilation lifetimes, employing the zero-positron-density limit of the two-component density functional theory [19, 37]. The first-principles electronic structure calculations of CdSe were performed using the PAW method [38] as implemented in VASP [39-41]. The plane wave cutoff energy was set to 357 eV in all calculations. Brillouin zone integration was performed using an 11×11×7 Γ-centered k-grid for the hexagonal bulk cell, and two-dimensional grids of a comparable density were used for the slab calculations. Electron-electron exchange and correlation was described with the Perdew-Burke-Ernzerhof functional [42]. Surfaces were modeled in a slab geometry with a vacuum region of at least 1.5 nm. The lattice parameters and atomic positions were optimized for the bulk unit cell prior to the construction of the slab models in which ionic positions were optimized but lattice parameters were kept fixed. We only considered non-polar surfaces of CdSe.

For the positron ground state calculations, we considered two models to describe the electron-positron correlation potential, namely, LDA+CMM, and WDA. In CMM [22], the erroneous LDA potential [43] in the vacuum regions of the simulation cell is empirically corrected to give the correct asymptotic $\sim 1/(z-z_0)$ behavior. For positron lifetime calculations, we set the enhancement factor $\gamma$ [44] to unity wherever the image potential was imposed, instead of using the LDA enhancement. Indeed, since the LDA enhancement factor implicitly assumes that the screening electron cloud is attached to the positron, its use is inconsistent with the use of the image potential since the screening cloud resides at the surface of the material [23]. In contrast, the WDA has the correct asymptotic behavior far away from the surface, and hence it does not require an empirical correction of the potential and the enhancement factor [21], but corrections are necessary in order to reproduce the experimental bulk lifetimes [24, 45, 46]. Here we apply the shell-partitioning method, treating the Cd(5s), Cd(4d), S(4s) and Se(4p) electrons as valence electrons, and a modified screening charge $Q$ [24], assumed to be equal to the charge of a single electron in earlier work [45, 46]. In this way, an adequate description of the screening can be obtained, which in turn can be expected to yield a good description of the electron-positron interaction potential at the surface. The details of our implementation of both models can be found in Refs. [24, 47].

Turning first to the LDA+CMM results (i.e. procedure of Ref. [47]), we obtained an image potential reference plane at $z_0$=1.8±0.1 a.u. (1 a.u.=0.0529 nm) from the topmost atom at the surface for both the $(10\bar{1}0)$ and $(11\bar{2}0)$ surfaces. Neither of the calculated positron lifetimes of respectively 251 ps and 257 ps are, however, near the range of the experimental values $\tau$=358–371 ps. In fact, these values are close to the LDA bulk value 246 ps, which is to be expected since the resulting positron state is seen to reside mostly inside the quantum dot (Figure 2(a)).

In the WDA calculations, we find that a modified screening charge of $Q$=1.35 reproduces the experimental *bulk* lifetime of 275 ps in CdSe [9]. Using this $Q$ value, we find a positron surface state at both considered surfaces, in sharp contrast to the LDA+CMM results, even though the tails of the states penetrate several layers into the material as seen in Figure 2(b). The computed energy difference of 0.18 eV between the shown state and the first bulk state for the CdSe ($10\bar{1}0$) slab confirms that these results indeed correspond to true surface states rather than surface resonances. The energy difference further shows that thermal excitation from the surface to the bulk state is negligible. The WDA-based lifetimes are much closer to the experimental values than the LDA+CMM, with values of 328 ps and 333 ps for the ($10\bar{1}0$) and ($11\bar{2}0$) surfaces, respectively, demonstrating that these computations provide a good description of the positron state at the surfaces of CdSe. It is highly satisfying that the $Q$ value fitted for the bulk provides a good description of the surface state and its positron lifetime value as well.

Moreover, the WDA provides a fundamental conceptual advance over the LDA+CMM scheme for the following reasons. (1) Unlike LDA+CMM, the asymptotic behavior of the potential in WDA away from the surface is not imposed by hand in an ad hoc manner, but it arises naturally through a proper description of the underlying electron-positron correlation cloud. (2) The WDA enhancement factor varies continuously from the bulk into the vacuum as it is obtained by the computed screening cloud, while in LDA+CMM it is abruptly replaced by unity when crossing the $z_0$ boundary into the vacuum region. And, (3) the $Q$ parameter in WDA involves a relatively simple bulk computation, while the $z_0$ parameter in LDA+CMM not only depends on the surface exposed but it is also difficult to generalize for surface geometries beyond a flat surface.

We finally discuss the sensitivity of the lifetimes and positron states to variation of $z_0$ and $Q$. Our purpose is to determine the parameter values which will reproduce exactly the measured lifetimes and examine the corresponding surface state. Figure 3(a) shows that in the LDA+CMM, in order to achieve consistency between the calculated and measured lifetimes, we need to shift the calculated $z_0$ to $z_0 \approx 2.4$ a.u. Figure 2(a) shows that this leads to strong localization of the positron state at the surface, overlapping significantly only with the few topmost Cd-Se layers. This demonstrates that within the LDA+CMM also the measured lifetimes indicate the presence of a surface state. Within the WDA, we can obtain exact agreement with the measured lifetime by a slight change of $Q=1.35$, which fits the bulk lifetime, to $Q=1.28-1.30$ (Fig. 3(b)). The slightly reduced $Q$ value at the surface likely reflects a decreased overlap of the positron with Cd(4d)-electrons, consistent with the experimental results of Ref. [10]. Note that the WDA predicts a significant penetration of the positron surface state into the bulk of the dots both for $Q=1.35$ and $Q=1.28-1.30$, as shown in Figure 2, indicating a qualitative difference with the LDA+CMM predictions at $z_0 \approx 2.4$ a.u.

Our analysis shows that LDA and LDA+CMM as well as the WDA are able to reproduce both bulk and surface lifetimes with a single adjustable parameter. However, LDA+CMM is unable to predict a surface state without having as input the corresponding experimental lifetime, while this is not the case for WDA, demonstrating its more advanced predictive capabilities. Finally, we note that the WDA has the potential to be quantitatively superior in applications that probe finer details of the positron-electron correlation and positron wavefunction.

Our in-depth PALS measurements of CdSe QDs demonstrate abundant trapping and annihilation of positrons at the surfaces of the QDs. Our parallel first-principles calculations within our WDA scheme confirm the existence of a positron surface state with calculated lifetimes that are close to the experimental values. Our work thus resolves the longstanding controversy concerning the nature of the positron state in QDs. We also demonstrate predictive capabilities of the WDA, and its conceptual superiority over conventional schemes such as the LDA+CMM, opening a new pathway for unraveling the complex behavior of positrons at the surface of QDs, which would allow us to obtain quantitative information on PEMDs of colloidal QDs. Our study thus provides a robust basis for the application of positron annihilation spectroscopy as a highly surface-sensitive tool for probing surface compositions and ligand-surface interactions of colloidal semiconductor QDs.

**Acknowledgments**

The work at Delft University of Technology was supported by the China Scholarship Council (CSC) grant of W.S. We acknowledge financial support for this research from ADEM, A green Deal in Energy Materials of the Ministry of Economic Affairs of The Netherlands (www.adem-innovationlab.nl). The PALS study is based upon experiments performed at the PLEPS instrument of the NEPOMUC facility at the Heinz Maier-Leibnitz Zentrum (MLZ), Garching, Germany, and was supported by the European Commission under the 7th Framework Programme, Key Action: Strengthening the European Research Area, Research Infrastructures, Contract No. 226507, NMI3. The work at the University of Maine was supported by the National Science Foundation under Grant No. DMR-1206940. V.C. and R.S. were supported by the FWO-Vlaanderen through Project No. G. 0224.14N. Computational resources and services used in this work were in part provided by the VSC (Flemish Supercomputer Center) and the HPC infrastructure of the University of Antwerp (CalcUA), both funded by the FWO-Vlaanderen and the Flemish Government (EWI Department). The work at Northeastern University was supported by the US Department of Energy (DOE), Office of Science, Basic Energy Sciences grant number DE-FG02-07ER46352 (core research), and benefited from Northeastern University's Advanced Scientific Computation Center (ASCC), the NERSC supercomputing center through DOE grant number DE-AC02-05CH11231, and support (functionals for modeling positron spectroscopies of layered materials) from the DOE EFRC: Center for the Computational Design of Functional Layered Materials (CCDM) under DE-SC0012575.

**Figures and captions**

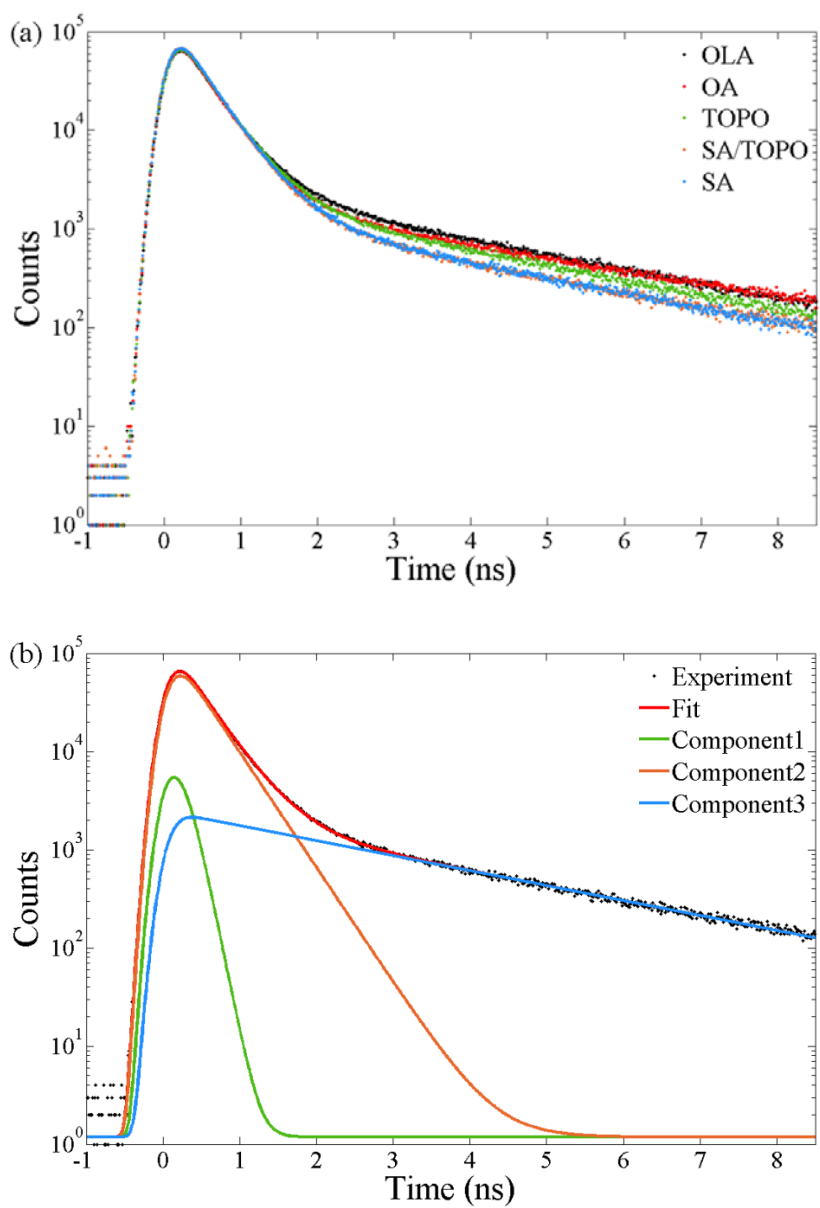

Figure 1: (a) Positron lifetime spectra of CdSe QDs capped with OLA, OA, TOPO, SA/TOPO, and SA ligands, collected at the positron implantation energy of 6 keV. (b) Data with TOPO ligands in (a) (solid circles) along with the corresponding fit (red full curve) and decomposition of the spectrum into three lifetime components (green, magenta, and blue lines) using the LT software.

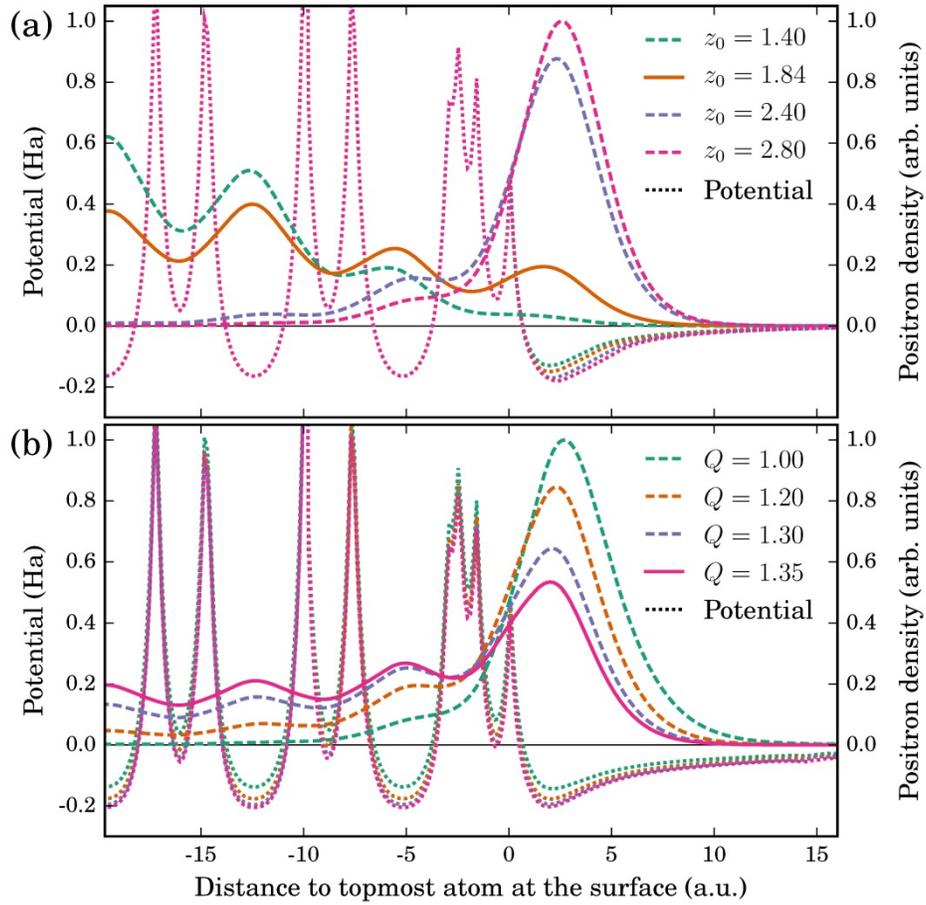

Figure 2: Results of the positron calculations at the CdSe $(10\bar{1}0)$ surface with (a) the LDA+CMM model, and (b) the WDA approach. Full curves show positron densities (normalized to the highest density in the plotted region) with $z_0$=1.837 a.u. and $Q$=1.35, determined from the background edge and the bulk lifetime, respectively. Dashed lines show the effect of variation of the parameters. Total potentials (Coulomb and correlation) are given by dotted lines. All curves give averages over the planes parallel to the surface.

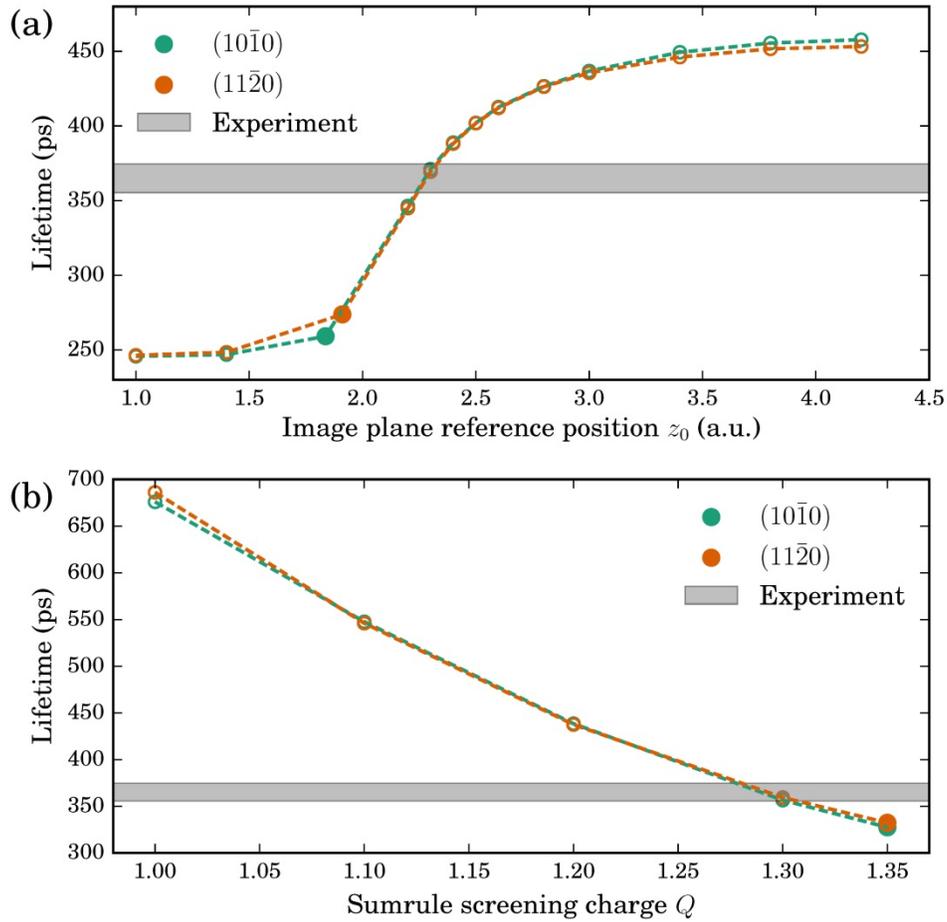

Figure 3: Computed positron lifetimes using (a) the LDA+CMM and (b) WDA for different values of the image potential reference plane $z_0$ and screening charge $Q$, respectively. Filled symbols indicate lifetimes calculated with $z_0$=1.837 a.u. determined from the background edge and $Q$=1.35 that fits the bulk lifetime. The gray area indicates the range of experimentally measured positron lifetimes at the surface.

*These two authors contributed equally to this work.

†Corresponding author: s.w.h.eijt@tudelft.nl  Phone: +31.15-278-9053

[1]     A. Stavrinadis, S. Pradhan, P. Papagiorgis, G. Itskos, and G. Konstantatos, ACS Energy Lett. **2**, 739 (2017).

[2]     M. A. Boles, D. Ling, T. Hyeon, and D. V. Talapin, Nature Mater. **15**, 141 (2016).

[3]     O. Voznyy and E.H. Sargent, Phys. Rev. Lett. **112**, 157401 (2014).

[4]     F.D. Ott, L.L. Spiegel, D.J. Norris, and S.C. Erwin, Phys. Rev. Lett. **113**, 156803 (2014).

[5]     A.H. Ip *et al*., Nature Nanotech. **7**, 577 (2012).

[6]     J.R.I. Lee, R.W. Meulenberg, K.M. Hanif, H. Mattoussi, J.E. Klepeis, L.J. Terminello, and T. van Buuren, Phys. Rev. Lett. **98**, 146803 (2007).

[7]     J. R.I. Lee *et al*., Nano Lett. **12**, 2763 (2012).

[8]     J. T. Wright and R. W. Meulenberg, Appl. Phys. Lett. **101**, 193104 (2012).

[9]     M.H. Weber, K.G. Lynn, B. Barbiellini, P.A. Sterne, and A.B. Denison, Phys. Rev. B **66** 041305 (2002).

[10]    S. W. H. Eijt, A. van Veen, H. Schut, P. E. Mijnarends, A. B. Denison, B. Barbiellini, and A. Bansil, Nature Mater. **5**, 23 (2006).

[11]    S.K. Sharma, K. Sudarshan, P. Maheshwari, D. Dutta, P.K. Pujari, C.P. Shah, M. Kumar, and P. Bajaj, Eur. Phys. J. B **82**, 335 (2011).

[12]    D. Kim, D.-H. Kim, J.-H. Lee, and J. C. Grossman, Phys. Rev. Lett. **110**, 196802 (2013).

[13]    A. Puzder, A.J. Williamson, F. Gygi, and G. Galli, Phys. Rev. Lett. **92**, 217401 (2004).

[14]    V. Petkov, I. Moreels, Z. Hens, and Y. Ren, Phys. Rev. B **81**, 241304 (2010).

[15]    A. Franceschetti, Phys. Rev. B **78**, 075418 (2008).

[16]    L. Chai *et al.*, APL Materials **1**, 022111 (2013).

[17]    W. Shi, S. W. H. Eijt, C. S. Suchand Sandeep, L. D. A. Siebbeles, A. J. Houtepen, S. Kinge, E. Brück, B. Barbiellini, and A. Bansil, Appl. Phys. Lett. **108**, 081602 (2016).

[18]    M. J. Puska and R. M. Nieminen. Rev. Mod. Phys. **66**, 841 (1994).

[19]    F. Tuomisto and I. Makkonen, Rev. Mod. Phys. **85**, 1583 (2013).

[20]    S. Mukherjee, M.P. Nadesalingam, P. Guagliardo, A.D. Sergeant, B. Barbiellini, J.F. Williams, N.G. Fazleev, and A.H. Weiss, Phys. Rev. Lett. **104**, 247403 (2010).

[21]    A. Rubaszek, Phys. Rev. B **44**, 10857 (1991).

[22]    R. M. Nieminen and M. J. Puska, Phys. Rev. Lett. **50**, 281 (1983).

[23]    R. M. Nieminen, M. J. Puska, and M. Manninen, Phys. Rev. Lett. **53**, 1298 (1984).

[24]    V. Callewaert, R. Saniz, B. Barbiellini, A. Bansil, and B. Partoens, Phys. Rev. B **96**, 085135 (2017).

[25]  V.A. Chirayath, V. Callewaert, A.J. Fairchild, M.D. Chrysler, R.W. Gladen, A.D. Mcdonald, S.K. Imam, K. Shastry, A.R. Koymen, R. Saniz, B. Barbiellini, K. Rajeshwar, B. Partoens, and A.H. Weiss, Nature Comm. **8**, 16116 (2017).

[26]  B. Shakeri and R. W. Meulenberg, Langmuir **31**, 13433 (2015).

[27]  R. Krause-Rehberg, and H. Leipner, *Positron Annihilation in Semiconductors - Defect Studies*, (Springer Verlag, Berlin, 1999).

[28]  P.J. Schultz, and K.G. Lynn, Rev. Mod. Phys. **60**, 701 (1988).

[29]  P. Sperr, W. Egger, G. Kögel, G. Dollinger, C. Hugenschmidt, R. Repper, and C. Piochacz, Appl. Surf. Sci. **255**, 35 (2008).

[30]  C. Hugenschmidt, B. Löwe, J. Mayer, C. Piochacz, P. Pikart, R. Repper, M. Stadlbauer, and K. Schreckenbach, Nucl. Instr. Meth. A **593**, 616 (2008).

[31]  D. Giebel, and J. Kansy, Phys. Proc. **35**, 122 (2012).

[32]  F. Plazaola, A.P. Seitsonen, M.J. Puska, J. Phys.: Condens. Matter **6**, 8809 (1994).

[33]  W. Egger, P. Sperr, G. Kögel, M. Wetzel, and H. J. Gudladt, Appl. Surf. Sci. **255**, 209 (2008).

[34]  K. G. Lynn, W. E. Frieze, and P. J. Schultz, Phys. Rev. Lett. **52**, 1137 (1984).

[35]  D. W. Gidley, H.-G. Peng, and R.S. Vallery, Ann. Rev. Mater. Res. **36**, 49 (2006).

[36]  P. Crivelli, U. Gendotti, A. Rubbia, L. Liszkay, P. Perez, and C. Corbel, Phys. Rev. A **81**, 052703 (2010).

[37]  E. Boroński and R. M. Nieminen, Phys. Rev. B **34**, 3820 (1986).

[38]  P. Blöchl, Phys. Rev. B **50**, 17953 (1994).

[39]  G. Kresse and J. Furthmüller, Comp. Mater. Sci. **6**, 15 (1996).

[40]  G. Kresse and J. Furthmüller, Phys. Rev. B **54**, 11169 (1996).

[41]  G. Kresse and D. Joubert, Phys. Rev. B **59**, 1758 (1999).

[42]  J. P. Perdew, K. Burke, and M. Ernzerhof, Phys. Rev. Lett. **77**, 3865 (1996).

[43]  N. D. Drummond, P. López Ríos, R. J. Needs and C. J. Pickard, Phys. Rev. Lett. **107**, 207402 (2011).

[44]  J. Laverock, T. D. Haynes, M. A. Alam, S. B. Dugdale, Phys. Rev. B **82**, 125127 (2010).

[45]  A. Rubaszek, Z. Szotek, and W. M. Temmerman, Phys. Rev. B **58**, 11285 (1998).

[46]  A. Rubaszek, Z. Szotek, and W. M. Temmerman, Acta Phys. Pol. A **95**, 652 (1999).

[47]  V. Callewaert, K. Shastry, R. Saniz, I. Makkonen, B. Barbiellini, B. A. Assaf, D. Heiman, J. S. Moodera, B. Partoens, A. Bansil, A. H. Weiss, Phys. Rev. B **94**, 115411 (2016).